\documentclass{article}
\usepackage{chicago}
\usepackage{latexsym}

\newcommand{\iec}{\mbox{i.\,e.\,}}

\newcommand{\egc}{\mbox{e.\,g.\,}}

\newcommand{\vctr}[1]{\ensuremath{\mathbf{ #1 }}}

\newcommand{\pbp}[2]{\ensuremath{\frac{\partial #1}{\partial #2}}}

\newcommand{\ket}[1]{\ensuremath{\left|  #1 \right\rangle}}

\newcommand{\op}[1]{\ensuremath{\widehat{\textsf{\ensuremath{#1}}}}}

\newcommand{\id}{\op{\mathsf{1}}}

\newcommand{\tr}{\textsf{Tr}}
\newcommand{\nrm}{\frac{1} {\sqrt{2} } }
\newcommand{\mc}[1]{\ensuremath{\mathcal{#1}}}
\newcommand{\dr}[1]{\ensuremath{\mathrm{d} #1\,}}
\newcommand{\vct}[2]{\ensuremath{\left( \begin{array}{c} #1 \\ #2 \end{array} \right)}}

\newcommand{\be}{\begin{equation}}
\newcommand{\ee}{\end{equation}}
\newcommand{\e}[1]{\mathrm{e}^{#1}}

\begin{document}

\title{QFT, Antimatter, and Symmetry}
\author{David Wallace}
\maketitle
\begin{abstract}
A systematic analysis is made of the relations between the symmetries of a classical field and the symmetries of the one-particle quantum system that results from quantizing that field in regimes where interactions are weak. The results are applied to gain a greater insight into the phenomenon of antimatter.
\end{abstract}

\section{Introduction}

Quantum mechanics comes with a $U(1)$ symmetry built in. All quantum-mechanical theories are formulated on a complex Hilbert space; all quantum-mechanical systems obey a Schr\"{o}dinger equation that is invariant under phase transformations.

Classical field theory does not have a $U(1)$ symmetry built in, but one can be put in by hand. That is, ``complex'' fields are just as valid as real fields in classical field theory: they are just ordered pairs of real fields. And if the transformation of that ordered pair which corresponds to ``complex multiplication'' is a symmetry of the field equations, then the classical field theory has a $U(1)$ symmetry.

When the two sorts of $U(1)$ symmetries are present in the same theory --- that is, when we try to quantize a classical complex field --- it seems that interesting things happen. The quantization process seems to have to be changed; the quanta which emerge seem to come in particle and antiparticle varieties; much confusion ensues.

Indeed, sometimes one can get the impression that the rules for quantizing a classical field are fundamentally different depending on whether that field is real or complex. Folk history --- and, to some extent, real history --- may seem to support this: applying the real-field quantization process to a complex field leads to the pathology of negative energy states, which (it can seem) can be removed only by the ad hoc reinterpretation of those solutions as antiparticles.

By contrast, my theme in this paper is that the same quantization process applies to real and complex fields, but that needless confusion ensues as long as we forget that the two $U(1)$ symmetries are \emph{distinct}.  One of them is a universal feature of all quantum theories; the other is specific to certain field theories, and in foundational issues as widely separated as the nature of the gauge principle and the origins of antimatter, it is vital that we make a careful distinction between the two.

In fact, to properly understand the quantization process and the origin of antimatter, it is necessary to develop a general theory of how classical symmetries behave under quantization, and the main content of this paper is exactly that. To be more precise: my main result is a systematic account of exactly what the relation is between the symmetries of a classical field theory and those of the corresponding quantum-mechanical particle. 

Nothing here is exactly new: in particular, the importance of distinguishing two forms of complex symmetry in QFT has been stressed by \citeN{saunderscomplexnumbers}, and of course the details of what happens when one or other classical field is quantized can be teased out of innumerable QFT textbooks. But there is, I think, some value in appreciating the systematic shape of quantization theory as it applies to symmetries. A great deal that can seem like black magic is thereby displayed as simple and natural. 

By way of motivation, I begin my account  (in sections \ref{sect2}--\ref{sect3}) by considering two puzzling aspects of modern quantum mechanics. In section \ref{sect2}, I sharply criticise the standard presentation of the gauge principle in quantum mechanics; in section \ref{sect3} I try to persuade the reader just why antimatter is prima facie so puzzling. 

In section \ref{linearclassical} I develop the classical theory of linear fields and their symmetries, as a precursor to the quantum theory of those fields which I present in section \ref{linearquantum}. Real field theories, of course, are not linear, and I sketch the connection between linear and nonlinear results in section \ref{nonlinear}. (My discussion takes place in Lagrangian QFT --- \iec, the sort used in theoretical particle physics and solid-state physics --- and is only partially applicable to axiomatic and algebraic approaches to QFT. See~\citeN{bakerhalvorson} for a discussion of antimatter from the algebraic perspective.)

Sections \ref{irreducibility} and \ref{generalise} are the heart of the paper. Here I set out the details of what the symmetries of the classical field theory do and do not entail for symmetries of the corresponding quantum-mechanical particles; in the process, I clarify just why quantizing complex fields leads to the antimatter phenomenon, and return to the question of the gauge principle. Sections \ref{irreducibility}--\ref{generalise} deal with small symmetries; \ref{CPT} extends the theory to parity, conjugation and time reversal symmetries. In section \ref{conclusion} I make a few concluding remarks.

At various points in the argument I make fairly extensive use of the theory of complexification (of vector spaces, and of representations upon those spaces). I sketch the details in the main part of the paper, but to avoid breaking up the argument I have relegated a full and careful development of these results to an  Appendix. In another appendix I apply the theory I have presented to the Standard Model of particle physics. (This material is relegated to an appendix since it is considerably more technical than the rest of the paper, and makes extensive use of results in QFT which space prevents me from explaining in detail).

\section{First preamble: the gauge principle in non-relativistic quantum mechanics}\label{sect2}

There is a standard textbook argument to motivate the gauge principle in quantum mechanics, which goes something like this. 
\begin{enumerate}
\item
Firstly, it is noted that global phase transformations of the wavefunction are not empirically detectable, and so $\psi(x)$ and $\mathrm{e}^{i \theta}\psi(x)$ are really just different representations of the same physical state.\footnote{Roughly, by a ``global'' transformation I mean one specified by a finite set of parameters; by a ``local'' transformation I mean one specified by a finite set of spacetime functions. But in any case, in this paper I make use of them only to motivate my analysis; they play no part in the analysis proper.}
\item Then it is argued that this should continue to be so if the phase transformation is performed not on the \emph{whole} wavefunction but only on a part of it --- or, equivalently, that $\psi(x)$ and $\mathrm{e}^{i \theta}\psi(x)$ should continue to be different mathematical representations of the same state \emph{even if $\theta$ is position-dependent}. The motivation for this move is normally that the probability of finding a particle at position $x$ is $|\psi(x)|^2$, a quantity which is invariant under even position-dependent phase transformations. 
\item It is then noted that the Schr\"{o}dinger equation is not invariant under such a transformation, so a new (classical) field, the $\vctr{A}$-field --- often called  the \emph{connection} --- is introduced to compensate for this. In more detail (recall), $\nabla$ is replaced with $(\nabla-i q \vctr{A})$, and a phase transformation is required to replace $\vctr{A}$ with $\vctr{A}-\nabla \theta$.  
\item Finally, a (classical) dynamics is introduced for \vctr{A}, given by a Lagrangian density whose form is not forced by the argument but is in fact taken to be the usual Lagrangian of vacuum electromagnetism.
\end{enumerate}

It's not a very good argument. It's long been recognised that the last step (in which \vctr{A} becomes a dynamical player)  needs further motivation: all that the first three steps motivate is the existence of a static, possibly even flat (\iec, satisfying $\nabla \times \vctr{A}=0$) connection (see, \egc, \citeNP{anandanbrown}). But in fact the problems begin earlier than this.

\begin{enumerate}
\item For a start, there is something rather awkward about the setting in which the argument is presented. We are dealing with a \emph{quantum} system (nonrelativistic particle mechanics), yet we are motivated to introduce a \emph{classical} connection (with associated classical dynamics). To be sure, we are used to the unfortunate process of constructing a theory classically (applying various arguments to justify its form) and then quantizing it, but something seems yet more unfortunate about constructing our theories at the part-quantized level. We could, of course, see the argument as occurring after first quantization (which gives us quantum particle mechanics) but before second quantization (which gives us quantum field theory, including a properly quantum theory of the connection). But it has long been recognised that ``second quantization'' is not a perspicuous way to understand quantum field theory, and that we do better to begin with a classical field and quantize it only once.
\item Consider the second step: going from a global to a local symmetry. This is motivated by the claim that \emph{position} measurements are invariant under local symmetries; but position measurements are not the only measurements that can be made, and more seriously, the squared modulus of the wavefunction is not the only physically relevant property it has. Why not apply the gauge principle to the momentum-space representation of the quantum state? That gives a perfectly well-defined physical theory, but one in which the ``magnetic field'' would be a field on momentum space rather than physical space, and so would be very different to the magnetic fields actually observed in nature.
\item Even supposing that position measurements really are preferred in a defensible sense, we still have problems. For the wavefunction, in general, is not defined on physical space at all: it is defined on $3N-$dimensional \emph{configuration space}. And if we perform a phase transformation where the phase is not only spatial-position-dependent but \emph{configuration-space-position-dependent}, again this will leave all probabilities of measurements of particle position unaffected. Again we will get a perfectly well-defined theory this way (with a connection on configuration space rather than on physical space); again this theory bears little resemblance to anything actually found in the world.
\item Changing tack: how do neutral particles fit into the argument? It is supposed to apply to the wavefunction of any one-particle system, yet neutral particles do not interact at all with the \vctr{A}-field. (I ignore magnetic effects here: they are in any case not predicted by the gauge argument).
\item Finally, the argument creates a worrying disanalogy between \emph{electromagnetic} gauge theory and other gauge theories (such as the SU(3) theory that is quantum chromodynamics (QCD) and the spontaneously broken SU(2) theory which contributes to the weak interaction). Superficially the arguments look similar: in QCD, for instance, particles have an internal (``colour'') degree of freedom, and global rotations of that degree of freedom are a symmetry of the system's dynamics. We can localise that symmetry if we introduce a (rather complicated) connection, an  su(3)-valued vector field,\footnote{That is, by a vector field whose components are elements of su(3) rather than real numbers (or, if a coordinate-free definition is desired, a field of linear maps from one-forms into su(3)).} and we can postulate a dynamics for it just as for the electromagnetic connection. 

But the SU(3) symmetry here is a very different object from the phase symmetry which grounds electromagnetism according to the standard argument. Only very special theories have an SU(3) internal symmetry, whereas \emph{every} quantum theory has global phase as a symmetry. In fact, calling it a ``symmetry'' at all is generous: it is natural, and common, to formulate QM in forms which eliminate it entirely (in terms of density operators, for instance, or of rays on Hilbert space.) Global  SU(3), by contrast, is a symmetry in the fullest sense, as much so as translation or rotation. (Note that SU(3) symmetry generates perfectly respectable conserved quantities; the conserved quantity associated with global phase symmetry, by contrast, is trivial).
\end{enumerate}

In fact, it seems to me that the standard argument feels convincing only because, when using it, we forget what the wave-function really is. It is not a complex classical  field on spacetime, yet the standard argument, in effect, assumes that it is. This in turn suggests that the true home of the gauge argument is not non-relativistic quantum mechanics, but classical field theory.

On the other hand, the standard argument \emph{works}. If the gauge principle is really understood at the field-theoretic level, presumably the electromagnetic interaction in non-relativistic quantum mechanics must be understood by thinking of that theory as a low-energy special case of quantum field theory. But then, what explains the success of the standard argument when applied \emph{directly} to non-relativistic quantum mechanics? ``Coincidence'' feels unsatisfactory.

\section{Second preamble: antimatter}\label{sect3}

So much for \emph{local} internal symmetries of quantum theory. A mystery of a rather different kind emerges when we consider just the \emph{global} symmetries. Consider: from a classical perspective a complex field is just a real field with an internal degree of freedom,\footnote{This is not \emph{quite} true, even classically. Massless spinor fields are irreducibly complex, and Lorentz transformations apply phase shifts to the field as well as rotating it on spacetime. But nothing similar happens for massive spinor fields (as the Majorana representation of the Dirac equation demonstrates), or for scalar and vector fields, so this phenomenon cannot be the root of the difference between the $U(1)$ internal symmetry and other internal symmetries.} not different in kind to, say, a field taking values in $\mathrm{R}^3$ or $\mathrm{C}^2$. In terms of symmetries, the complex field just has an internal $U(1)$ symmetry, whereas the field taking values in $\mathrm{R}^3$ has an $SU(3)$ internal symmetry and the field taking values in $\mathrm{C}^2$ has a $U(2)$ internal symmetry.

What happens when we quantize theories with internal symmetries like these? In general, things proceed rather as we would expect: the particles that emerge from the quantization have an internal degree of freedom. So quarks, for instance, have a three-dimensional internal degree of freedom which we call colour, resulting from the SU(3) symmetry of the quark field. It is convenient to pick a basis in that internal space and label its vectors \emph{red, green, blue}, but that choice of basis is arbitrary --- we could just as easily have chosen \emph{$\nrm$(red+green), $\nrm$(red-green), blue}. We might expect, then, that something similar would happen when we quantize a theory with a $U(1)$ symmetry: that is, we might expect to find such theories have a two-dimensional internal degree of freedom, with no particularly preferred basis in it. 

If we did expect that, we would be astonished. What actually happens, of course, is that quantizing a complex field produces \emph{antimatter}. To be sure, the existence of matter and antimatter versions of the quantized field's particles is an additional degree of freedom, but it behaves altogether differently from the SU(3) case. There is nothing at all arbitrary about our preference for the \emph{matter, antimatter} basis over the \emph{$\nrm$(matter + antimatter), $\nrm$(matter - antimatter)} basis, and nothing at all arbitrary about saying that our world is made up of one of the two kinds of particles rather than of a linear superposition of the two kinds.\footnote{The reader who doubts this is invited to perform the following experiment, preferably by remote control and far away from any population centre: prepare a few kilograms of particles which are either (a) all definitely matter or all definitely antimatter, or (b) all definitely in a matter-antimatter superposition; isolate them from their environment; wait for a few milliseconds. If any living thing remains within a quarter-mile of your laboratory, you have case (a). (At the mathematical level, if each individual atom is in a superposition $(1/\sqrt{2})(\ket{\mbox{matter}}+\ket{\mbox{antimatter}})$, the tensor product of such states for a macroscopic number $N$ of such atoms can be expanded into states containing $M$ antimatter atoms and $N-M$ matter atoms; basic combinatorics tells us that the expansion is dominated by terms containing $\sim N/2$ of each, and quantum electrodynamics tells us that the matter particles will quickly annihilate the antimatter particles in each superposition.)} 

So we have a puzzle: why does quantizing a complex field not give rise to particles with an internal symmetry? (Or, if you prefer: why does quantizing fields with other internal symmetries not give rise to some generalisation of antimatter?)

Hopefully, you should by now have the impression --- both from antimatter, and from the gauge argument --- that something odd and interesting  happens in the quantization of complex fields. We now turn to the main business of the paper: establishing just how that quantization process works.

\section{Linear classical fields and their symmetries}\label{linearclassical}

We begin our analysis with linear field theories, partly because in the classical context they are the simplest field theories there are, and partly because of the well known duality between a classical linear field theory and a one-particle quantum theory. Any such theory can be specified by:
\begin{itemize}
\item A semi-Riemannian manifold \mc{M}, representing spacetime. For simplicity, I will always take \mc{M} to be \emph{Minkowski} spacetime, although most of the results generalise to other spacetimes.
\item A real or complex vector space \mc{V}, and an associated \mc{V}-bundle (that is, vector bundle with typical fibre \mc{V}) over \mc{M}, whose sections represent kinematically possible fields. (The technical details of vector bundles play little role in what follows: the reader can, if desired, safely replace ``section of a \mc{V}-bundle over \mc{M}'' with ``smooth function from \mc{M} to \mc{V}.'')
\item A Lagrangian density, quadratic in the fields and their first derivatives, which determines which kinematically possible fields are dynamically possible via the Euler-Lagrange equations.
\end{itemize}

Examples include:
\begin{description}
\item[A. Real Klein-Gordon theory:] Here the vector space is just the real line, so that fields are smooth real-valued functions on \mc{M}. The Lagrangian density is 
\be
\mc{L}=\frac{1}{2}(\partial_\mu \varphi \partial^\mu \varphi + m^2 \varphi^2)
\ee
and the associated equation of motion is
\be
(\partial^\mu \partial_\mu + m^2)\varphi=0.
\ee
\item[B. Complex Klein-Gordon theory:] Here instead we take fields to be smooth \emph{complex-valued} functions on \mc{M}, use the Lagrangian density
\be
\mc{L}=\frac{1}{2}(\partial_\mu \varphi^* \partial^\mu \varphi + m^2 \varphi^*\varphi)
\ee
and obtain the same equation of motion as in the real case.
\item[C. Klein-Gordon theory with internal degrees of freedom:] Here we take \mc{V} to be any finite-dimensional real vector space (whose vectors we write $v^a$) on which $h_{ab}$ is a positive-definite metric. The fields are now \mc{V}-valued functions, the Lagrangian is
\be
\mc{L}=\frac{1}{2}(h_{ab}(\partial_\mu \varphi^a \partial^\mu \varphi^b) + m^2 h_{ab}\varphi^a\varphi^b).
\ee
and the equation of motion is 
\be
(\partial^\mu \partial_\mu + m^2)\varphi^a=0.
\ee
(The further generalisation to complex Klein-Gordon theory with internal degrees of freedom is straightforward.)
\item[D. Weyl spinor theory:] Here \mc{V} is a two-dimensional complex vector space (more precisely, the vector bundle for the theory is the spin bundle over \mc{M}) on which $\epsilon_{ab}$ is a completely antisymmetric 2-tensor. Fields can then be thought of as ordered pairs of complex functions on \mc{M}, and we write them as $\phi^a$. One possible form of the Lagrangian density is now 
\be
\mc{L}=\epsilon_{ab}\phi^a(\partial_0+\sigma^i \partial_i)\phi^b
\ee
(where $\sigma^1,\sigma^2,\sigma^3$ are the Pauli matrices)
and the equation of motion is in any case
\be
(\partial_0+\sigma^i \partial_i)\phi^b=0.
\ee
Alternatively, if we take the Lagrangian to be 
\be
\mc{L}=\epsilon_{ab}\phi^a(\partial_0-\sigma^i \partial_i)\phi^b
\ee
then the equations of motion are
\be
(\partial_0-\sigma^i \partial_i)\phi^b=0.
\ee
So there are actually two kinds of Weyl fields, referred to as ``left-handed'' and ``right-handed'' in recognition of the fact that one is a mirror image of the other.
\item[E. Dirac spinor theory:] Dirac theory can be written in a number of ways, but for our purposes a convenient way is to use the bispinor formalism: take \mc{V} as the direct sum of two two-dimensional complex vector spaces. A field is then an ordered pair $(\phi^a,\chi^b)$ of complex 2-vector fields. I omit the detailed form of the Lagrangian and simply give the dynamical equations:
\be
(\partial_0+\sigma^i \partial_i)\phi^b-im \chi^b=0;\,\,\,\,
(\partial_0-\sigma^i \partial_i)\chi^b+im \phi^b=0.
\ee 
In effect, a Dirac field is a left-handed and a right-handed Weyl field, coupled together so as to restore mirror symmetry.
\item[F. Majorana spinor theory:] If we impose the condition $\chi^a=i\sigma^2 \phi^{*a}$ on the Dirac spinor, we obtain the single equation
\be
(\partial_0+\sigma^i \partial_i)\phi^b+m\sigma^2 \phi^{*b}=0.
\ee
\item[G. Real vector theory:] Here \mc{V} is a copy of Minkowski spacetime (more precisely, the vector bundle for the theory is the tangent bundle over \mc{M}). Fields are then vector fields $A^\mu$ in the usual sense of that term; the Lagrangian is 
\be
\frac{1}{4}(\partial_\mu A_\nu-\partial_\nu A_\mu)(\partial^\mu A^\nu-\partial^\nu A^\mu)+\frac{1}{2}m^2 A_\mu A^\mu
\ee
and the equations of motion are
\be
\partial_\nu A^\nu=0;(\partial^\mu \partial_\mu + m^2)A^\nu=0.
\ee
Replacing the Lagrangian with 
\be
\frac{1}{4}(\partial_\mu A^*_\nu-\partial_\nu A^*_\mu)(\partial^\mu A^\nu-\partial^\nu A^\mu)+\frac{1}{2}m^2 A^*_\mu A^\mu
\ee
generalises this theory to a complex vector theory.
\item[H. Free gauge boson theory:] Here $\mc{V}=\mc{V}_{vect}\otimes \mathbf{g}$, where $\mc{V}_{vect}$ is a copy of Minkowski spacetime and $\mathbf{g}$ is the Lie algebra of the real, finite-dimensional Lie group \mc{G} (more precisely, the vector bundle for the theory is the tensor product of the tangent bundle over \mc{M} with the trivial $\mathbf{g}$-bundle over $\mc{M}$). Fields are then $\mathbf{g}$-valued vector fields $A^\mu$; the Lagrangian is 
\be
\frac{1}{4}\tr\{(\partial_\mu A_\nu-\partial_\nu A_\mu)(\partial^\mu A^\nu-\partial^\nu A^\mu)\}
\ee
and the equations of motion are
\be
\partial^\mu \partial_\nu A^\nu- \partial_\nu \partial^\nu A^\mu=0.
\ee
\end{description}
Further generalisations are possible (albeit rarely physically relevant), and internal degrees of freedom can be added for vector and spinor fields in the same manner as for scalar fields. It should be noted that, in all these cases, the components of the various solutions always obey the Klein-Gordon equation.

So: it appears that free fields come in real and complex varieties, and also come with and without internal degrees of freedom. But this paper will take a somewhat different perspective: namely, \emph{complex fields are just special cases of real fields}. After all, any complex vector space is also a real vector space (of twice the dimension); any complex-linear theory is also a real-linear theory. Indeed, from this perspective, complex Klein-Gordon theory is just a special case of real Klein-Gordon theory with an internal degree of freedom. 

To elaborate: a complex vector space is, by definition, an additive group of vectors together with a rule for multiplying vectors by complex numbers such that the multiplication rule obeys certain constraints. That rule can be broken down into two parts: a rule for multiplying vectors by \emph{real} numbers, together with a real-linear map from vectors to vectors which represents multiplication by $i$. If we write that map as \textbf{J}, we have that
\be
(\alpha + i \beta)\vctr{v}=\alpha \vctr{v}+\beta \textbf{J}\vctr{v}.
\ee
And in fact, if \mc{V} is \emph{any} real vector space with a real-linear map $\mathbf{J}:\mc{V}\rightarrow\mc{V}$ satisfying $\mathbf{J}^2=-1$, this equality suffices to make \mc{V} into a complex vector space; for this reason, any such map is known as a \emph{complex structure} (a more careful mathematical treatment may be found in the Appendix).

So no true generalisation of real linear field theory is gained by allowing complex fields as a separate case. Rather: a complex linear field is just a real-linear field defined on a \mc{V}-bundle such that (i) there is a complex structure \textbf{J} on \mc{V}, and (ii) if $\varphi$ is a solution to the equations of motion, so is $\mathbf{J}\varphi$.
 
Of course, (ii) is just another way of saying that multiplication by \textbf{J} is a symmetry of the theory, and this brings us on to the more general question of what the symmetries of a field theory are. For our purposes, a symmetry may be taken to be any smooth fibre-preserving map $f$ of the field bundle onto itself which takes dynamically possible fields to other dynamically possible fields. (Recall that a fibre-preserving map is one where two vectors initially in the same fibre are not taken to different fibres, so that $f$ induces a diffeomorphism $\overline f$ on \mc{M}; in less geometric language, a fibre-preserving map is a map $\overline f$ of the spacetime \mc{M} onto itself, together with a rule, for each spacetime point $x$, taking vectors at $x$ to vectors at $\overline f(x)$).

We will, however, be interested mainly in \emph{Lagrangian} symmetries: symmetries with the property that the Lagrangian density, evaluated at $\overline f(x)$ with respect to the new field $f\cdot\phi$, is equal to the Lagrangian density evaluated at $x$ with respect to the old field $\phi$. These in turn can usefully be divided into three categories:
\begin{enumerate}
\item \emph{Rigid internal symmetries}, where $f\cdot \phi(x)$ depends only on $\phi(x)$: all such symmetries can be written as
\be
f \cdot \phi(x)=U(\phi(x))
\ee 
for some fixed $U:\mc{V}\rightarrow \mc{V}$.
\item \emph{Gauge internal symmetries}, where $f\cdot \phi(x)$ depends on $\phi(x)$ and $x$: all such symmetries can be written as
\be
f \cdot \phi(x)=U(x)(\phi(x))
\ee 
where $U(x)$ is some spacetime-dependent map from $\mc{V}$ to itself.
\item \emph{Spacetime symmetries}, where $f\cdot\phi(x)$ is not determined by $x$ and $\phi(x)$ alone.
\end{enumerate}
(My distinction between `rigid' and `gauge' symmetries is basically the same as the distinction between `global' and `local'; I resort to neologisms instead of using the standard terminology in order to avoid getting bogged down in semantic questions as to whether this is the ``right'' definition of a local symmetry.)
As is well known, the symmetries of a theory form a group; clearly, the rigid internal symmetries form a subgroup of that group.
As usual, we also distinguish between ``small'' and ``large'' symmetries: the former, but not the latter, lie in the connected component of the symmetry group containing the identity.
Spacetime symmetries are usually generated by some underlying symmetry of the spacetime metric: in the examples above,  then the familiar (small) Poincar\'{e} symmetries are spacetime symmetries. My concern will mostly be with internal symmetries, and indeed mostly with \emph{rigid} internal symmetries. For brevity, in fact, I will drop the terms ``small'' and ``rigid'', so that an internal symmetry is small and rigid unless otherwise stated. In particular, I use ``internal symmetry group'' to denote that connected component of the group of rigid symmetries which contains the identity.

In the cases considered above, for instance:
\begin{itemize}
\item In cases $A$, $G$ and $F$ (real scalar and vector fields, and the Majorana spinor field), the internal symmetry group is trivial.
\item In cases $B$, $D$ and $E$  (complex scalar fields, and Weyl and Dirac spinor fields), the internal symmetry group is $U(1)$: the action of $\cos \theta \id + \sin \theta \mathbf{J}$ is  a symmetry for any value of $\theta$, and $\theta$ generates the same transformation as $\theta + 2N \pi$.
\item In case $C$ (scalar fields with internal degrees of freedom), the internal symmetry group is $SO(N)$. (Note that, as a special case, when $N=2$ then the internal symmetry group is $SO(2)\simeq U(1)$: as promised, complex scalar field theory is a special case of $C$).
\item In case $H$ (free gauge boson theory), the internal symmetry group is $SO(Dim(g))$, where $Dim(g)$ is the dimension of the Lie algebra. Notice that this may not coincide with (the origin-containing component of) \mc{G}, the Lie group of which \textbf{g} is the Lie algebra, as a subgroup. If $\mc{G}=SU(2)$, for instance, the internal symmetry group is $SO(3)$, which at least is locally isomorphic to $SU(2)$; if $\mc{G}=SU(3)$, then the internal symmetry group is $SO(8)$, which is a considerably larger group.
\end{itemize}
In the rest of this paper, I will need some further assumptions about the symmetry groups of the fields I consider. Firstly, I will assume that the internal symmetry group is compact (all physical fields seem to have this property). As is well known, this entails the existence of an inner product on \mc{V} invariant under internal symmetry transformations.\footnote{To construct this inner product explicitly, let $\{v,w\}$ be an arbitrary inner product on \mc{V}, and define
\[
(v,w)=\int_\mc{G}\dr{\mu}(g)\{g\cdot v,g \cdot w\},
\]
where \mc{G} is the internal symmetry group, $g\cdot v$ is the action of $g\in \mc{G}$ on the vector $g$, and $\mu$ is the Haar measure.}

Secondly, I will assume that the internal symmetries (rigid and gauge) commute with the spacetime symmetries. (In the absence of supersymmetry, the Coleman-Mandula theorem proves that this must be the case; see, \egc, \citeN{wessbagger}.)

Finally, I will assume that the internal symmetry group acts \emph{irreducibly} on \mc{V}. In a sense this is no restriction at all: if the internal group acts reducibly (so that $\mc{V}=\mc{V}_1 \oplus \mc{V}_2$, and each $\mc{V}_i$ is invariant under the action of $g$) then we could perfectly well regard the theory as two fields instead of one, with one field taking values in $\mc{V}_1$ and the other in $\mc{V}_2$. One shallow reason for the requirement, then, is just that it is in some sense more ``natural'' to regard reducible theories as multi-field theories\footnote{This was basically the motivation adopted by Wigner in his classic~\citeyear{wignerclassification} classification of quantum \emph{particles} in terms of \emph{irreducible} representations of the Poincar\'{e} group.}; in the quantum case, renormalisation offers a deeper reason, as will be seen in section \ref{nonlinear}.

\section{Quantizing linear field theories}\label{linearquantum}

We are interested in extracting the properties of particles from a quantized
field theory, and particle phenomenology is associated with a free field theory
--- that is, with a field theory with linear dynamical equations. According
to the conventional position in theoretical particle physics (which I refer to hereafter as \emph{Lagrangian QFT}), this is not because of any
pathology of interacting field theories, but simply because particles are a useful
approximation, appropriate only in regimes where the interactions are relatively
small and can be treated perturbatively. According to various more formal approaches to QFT
 (notably, algebraic quantum field theory), it is because we do
not understand how to quantize interacting field theories at all, the success of
Lagrangian QFT notwithstanding.

The Lagrangian position will come to the forefront in the next section, but for now, we can confine our attention
to the free-field case. We will also assume that the theory has no gauge symmetries (such symmetries are generally dealt with by adopting some gauge-fixing convention prior to quantization, though there are ways which preserve the gauge symmetry at the quantum level). 
Because these theories have linear field equations, the solutions of these equations can be expressed as a sum of so-called \emph{normal-mode} solutions: solutions of the form
\be
\phi(x,t)=f_k(x)(C_k \exp(-i \omega_k t)+C_k^* \exp(+i \omega_k t))
\ee
for some function $f_k$ (which may  take values in the internal space of the theory, for instance by being spinor-, vector-, or su(3)-valued), some complex number $C_k$, and some positive real number $\omega_k$ (for details, see any text on field theory). A general solution to the equations can then be written as 
\be
\phi(x,t)=\int \dr{k}f_k(x)(C_k \exp(-i \omega_k t))+f_k^*(x)(C_k^* \exp(+i \omega_k t))
\ee
where the ``integral'' over $k$ is schematic, and might include discrete sums and/or continuous integrals. 

Quantizing these theories, in outline, is like quantizing any theory: the end product should be a Hilbert space (call it \mc{F}) representing the possible states of the field theory, together with various operators $\op{\psi}(x,t)$, $\partial_\mu \op{\psi}(x,t)$ which are the quantizations of the classical observables. (Since these field theories have internal degrees of freedom, their operators will have indices which range over those internal degrees of freedom.)

The quantization can be performed in a variety of ways, but the outcome is the same: the Hilbert space of the quantum field theory is the (symmetric or antisymmetric) \emph{Fock space}
\be
\mc{F}=\sum_{N=0}^\infty \mc{S}_N\mc{H}^N_{1P}.
\ee
Here:
\begin{enumerate}
\item $\mc{S}_N$ is the $N$-fold symmetrisation or antisymmetrisation operator;
\item $\mc{H}^N$ is the $N$-fold tensor product of $\mc{H}$ (so $\mc{H}^3=\mc{H}\otimes \mc{H}\otimes \mc{H}$, for instance);
\item $\mc{H}_{1P}$ is the \emph{one-particle Hilbert space}. 
\end{enumerate}
The symmetric case, of course, corresponds to bosons; the antisymmetric case, to fermions. So the problem of how to quantize a free field theory reduces to two questions: whether the field is bosonic or fermionic, and what the one-particle subspace is. My concern in this paper is entirely with the latter.

The one-particle subspace can be constructed from the space of solutions to the classical equations of motion in a fairly algorithmic way (again, the algorithm can be derived in a variety of ways). We  begin with the real-linear space $\mc{S}$ of classical solutions to the field equations; recall that each element of \mc{S} is a section of a \mc{V}-bundle over \mc{M}. Then we complexify \mc{S}: that is, replace it with the space $\mc{S}^\mc{C}=\mc{S}\oplus \mc{S}$, equipped with the complex structure \textbf{J} defined by $\mathbf{J}(v,w)=(-w,v)$. This is equivalent to complexifying the \mc{V}-bundle (to produce a bundle whose typical fibre is the complexification $\mc{V}^\mc{C}$ of \mc{V}) and taking $\mc{S}^\mc{C}$ to be those sections of this bundle which satisfy the dynamical equations. (Again, a more careful mathematical discussion of complexification can be found in the Appendix).

Next, we fix a foliation of \mc{M} by hyperplanes, and a direction of increasing time along that foliation. This defines a time coordinate $t$, and any element of \mc{S} can thus be written as $\psi(\vctr{x},t)$, where $\psi(\vctr{x},t)$ is a vector in \mc{V}. We can then construct the Fourier transform $\hat\psi(x,\omega)$ of $\psi(x,t)$ with respect to $t$, and thus divide any solution into its positive and negative frequency parts: $\psi=\psi_+ +\psi_-$, where $\hat\psi_+(x,\omega)$ is non-zero only for $\omega>0$ and $\hat\psi_-(x,\omega)$ is non-zero only for $\omega<0$. This process is uniquely defined and linear, so it divides $\mc{S}^\mc{C}$ into positive and negative frequency subspaces $\mc{S}^\mc{C}_+$ and $\mc{S}^\mc{C}_-$. We discard the negative-frequency subspace and work only with the positive-frequency one.
In terms of the modal analysis above, the positive-frequency solutions are the solutions of the form
\be\phi(x,t)=\int \dr{k}f_k(x)C_k \exp(-i \omega_k t).
\ee
 
Next, we provide $\mc{S}^\mc{C}_+$ with an inner product, which we do in two steps. Firstly, if $f$ and $g$ are positive-frequency solutions to the complex Klein-Gordon equation, we define 
\be
(f,g)=\int \dr{^3k} \frac{1}{\omega(k)}\hat f^*(k,\omega(k))\hat g(k,\omega(k))
\ee
where now $\hat f(k,k_0)$ is the Fourier transform of $f(x,t)$ with respect to both $x$ and $t$, and where $\omega(k)=+\sqrt{k\cdot k+m^2}$. This inner product is well known to be invariant under (small) Poincar\'{e} transformations.
And secondly, if $h_{ab}$ is a (real) inner product on $\mc{V}$ invariant under the action of the internal symmetry group (recall that the existence of such an inner product is entailed by the compactness of the internal symmetry group), we can define
\be
\langle\varphi,\psi\rangle=h_{ab}(\varphi^{a*},\psi^b).
\ee
for any $\varphi,\psi\in \mc{S}^\mc{C}_+$.
The result is a (complex) inner product for the complex vector space $\mc{S}^\mc{C}_+$, invariant under both internal and spacetime symmetries. Lastly, we turn $\mc{S}^\mc{C}_+$ into a Hilbert space by completing in this norm; the resultant Hilbert space is the one-particle space.

The functions in this space at least look like conventional one-particle wave-functions. In particular, in the trivial case of a scalar field with no internal degrees of freedom, they are simply complex functions on \mc{M} --- at any given time, they are just square-integrable complex functions on $\Sigma$. In the more general case of a field with internal degrees of freedom, they are maps from \mc{M} to the complexification $\mc{V}^C$ of \mc{V}. 
It is a somewhat complicated business to explain in exactly what sense these functions really \emph{can} be treated as wave-functions, but the bottom line is that actual physical practice is indeed to treat them as wave-functions, and that this practice appears to be justifiable. (See, \egc, chapters 1--3 of~\citeN{waldQFT} for the details, and~\citeN{wallaceqftloc} for an extended discussion.) 

For our purposes, the crucial result is this:
\begin{quote}
Single particles of a quantized free field obey the complexified version of the original field's dynamical equations, with the extra restriction that they must be positive-frequency, and the Hilbert space norm for their wavefunctions is invariant under symmetry transformations. 
\end{quote}

Unsurprisingly, there is a close relation between the symmetries of a classical field and the symmetries of the associated one-particle quantum theory. Since:
\begin{enumerate}
\item any classical symmetry leaves the dynamical equations of the classical linear field theory invariant;
\item any real transformation which leaves the dynamical equations invariant will also leave their complexification invariant;
\item the internal symmetries of the one-particle Hilbert space are exactly those Hilbert-space-norm-preserving transformations which leave the one-particle dynamical equations invariant;
\item the one-particle dynamical equations are just the complexification of the classical dynamical equations
\end{enumerate}
it follows that a classical symmetry is a symmetry of the one-particle subspace iff it takes positive-frequency solutions to positive-frequency solutions. In particular, the (small) Poincar\'{e} symmetries have this property; so do the internal symmetries.

It is natural to define internal symmetries of the one-particle theory in exactly the same way as for the classical theory. It follows that the internal symmetry \emph{group} of the one-particle theory is the same as for the classical theory, and the \emph{action} of that group is the \emph{complexification} of the action of the group on the classical theory (that is, the extension, by complex linearity, of the action on the real space of classical solutions to the complex space of positive-frequency solutions).

\section{Nonlinear field theories}\label{nonlinear}

If all fields were linear, the world would be boring: it is through nonlinearity that interactions enter physics. The true significance of linear field theory is that (i) in some circumstances (such as interactions with an external potential) it is a sufficient approximation to nonlinear theory in its own right; (ii) more importantly, in many circumstances we can treat the non-linear part of a theory as a perturbation of the linear part. The full details of how this works in quantum field theory are well beyond the scope of this paper (and of limited relevance to its goals); however, some important insights can be gained from a semi-classical discussion of the process. For more details, see the appropriate sections of, \egc,~\citeN{peskinschroeder}, \citeN{chengli}, or \citeN{coleman}.

It will be helpful to begin with a simple example: consider the Lagrangian
\be
\mathrm{L}=\frac{1}{2}\dot x^2-V(x)
\ee 
for a single one-dimensional particle, and suppose that $V$ is a smooth function with a global minimum at $x_0$. Then by elementary Taylor expansion, the Lagrangian can be written as
\be
\mathrm{L}=\frac{1}{2}\dot x^2-\frac{1}{2}V''(x_0)x^2 + \delta V(x-x_0)-V(x_0)
\ee
where $\delta V(x)$ is $o(x^2)$ (that is, $\delta V(x)/x^2 \rightarrow 0$ as $x \rightarrow 0$). In other words, provided that we are interested in motion sufficiently close to $x_0$ --- where ``sufficiently close'' will depend on the precise form of $V$ --- the Lagrangian of the theory is close to a simple harmonic oscillator, oscillating around $x_0$. If we quantize the theory, we would expect --- at least for a certain set of states --- that the theory can be treated as a harmonic oscillator together with a small correction term which can be analysed via perturbation theory.

Since a (bosonic) free-field quantum theory is --- mathematically speaking --- just a collection of harmonic oscillators, particles in quantum field theory can be understood in much the same way. Consider, for instance, the real Klein-Gordon theory  with the quadratic mass term  replaced by a more general potential:
\be
\mc{L}=\frac{1}{2}(\partial_\mu \varphi \partial^\mu \varphi)+ V(\varphi).
\ee
If $V$ has a minimum at $\varphi_0$, we can make the coordinate transformation $\varphi\rightarrow\varphi-\varphi_0$, and rewrite the Lagrangian
as
\be
\mc{L}=\frac{1}{2}(\partial_\mu \varphi \partial^\mu \varphi +V(\varphi_0)+ V''(\varphi_0)(\varphi-\varphi_0)^2)+\delta V (\varphi-\varphi_0),
\ee
where again $\delta V(\varphi)$ tends to zero at least as fast as $\varphi^3$ as $\varphi$ tends to zero. This theory now has the form of the Klein-Gordon equation (with $m=\sqrt{V''(\varphi_0)}$) together with a dynamically irrelevant constant term and a  perturbation term $\delta V(\varphi)$; prima facie we might expect that the perturbation can be treated as small, so that the theory can be analysed perturbatively as a free-field theory --- that is, a many-particle theory --- together with an interaction term which can be understood as generating scattering between particles.

This expectation is naive, though. In fact, the perturbation does not normally generate \emph{small} terms: it generates \emph{infinite} terms. This is the notorious ``problem of infinities'' of quantum field theory: the second-order and higher-order terms in the perturbative expansion for interacting quantum field theories, calculated formally, are infinitely large. The term ``problem'', however, is a misnomer (at least from the point of view of Langrangian QFT): the difficulty can be resolved in two steps. Firstly, the \emph{infinities} need to be understood as a consequence of naively assuming that the field theory can be defined for arbitrarily short lengthscales. If some sort of short-distance cutoff is imposed (most crudely, by replacing the continuum of spacetime points with a lattice) then the higher-order terms in the perturbative expansion become finite (albeit very large). This raises the problem that the dynamics become very sensitive to the details of the cutoff; in fact, though, it turns out that the effects of those details can be absorbed into adjustments to a very few parameters (basically the mass, the overall magnitude of the fields, and a small number of parameters determining the interaction term).  For instance,the Lagrangian density of the scalar theory above can be rewritten  as
\be
\mc{L}=\frac{1}{2}(\partial_\mu \varphi_R \partial^\mu \varphi_R + m_R^2\varphi_R^2)+\delta V_R (\varphi_R)+\mbox{constant},
\ee
where $\delta V_R$ really is ``small'' in perturbation-theory terms.\footnote{In fact, the form of $\delta V_R$ is sharply constrained by the renormalisation process: if we expand it as a power series, for instance, all but the $\varphi_R^3$ and $\varphi_R^4$ terms will have vanishingly small dynamical influence except at energy scales close to the cutoff scale; see \citeN{binney} or \citeN{peskinschroeder} for details. This plays no further role in my analysis, however.}
Of course, since the ``bare'' --- \iec, pre-adjustment --- values of these parameters are not experimentally accessible, what we actually measure is the renormalised parameters. Technical details of the renormalisation process can be found in, \egc, \citeN[pp.\,315--346]{peskinschroeder}, \citeN[pp.\,31--66]{chengli}, \citeN[pp.\,353--374]{binney} or \citeN[pp.\,99--112]{coleman}; for a more detailed conceptual discussion, see \citeN{wallaceconceptualqft}.

For the purposes of this paper, it is crucial to ask how symmetries of the full theory translate into symmetries of the linearised theory. At first sight, it might appear that \emph{any} symmetry of the former would be a symmetry of the latter; however, this fails to take into account the possibility that the point of expansion for the linear theory is itself not invariant under a symmetry transformation. This is the famous phenomenon of \emph{spontaneous symmetry breaking}: a classic example is the following special case\footnote{Given renormalisation, it actually isn't a particularly special case! --- this is the most general possible form of a renormalised complex scalar field theory.} of the scalar theory with internal degrees of freedom (case C in my earlier taxonomy):
\be
\mc{L}=\frac{1}{2}\langle\partial_\mu \varphi^, \partial^\mu \varphi\rangle+\alpha \langle\varphi,\varphi\rangle+\beta \langle\varphi,\varphi\rangle^2,
\ee
where $\alpha$ and $\beta$ are real numbers with $\beta>0$ and $\langle\cdot,\cdot\rangle$ is an inner product for the internal space \mc{V}.
Elementary calculus tells us that the location of the minimum of the constant-potential (that is, constant-$\varphi$) part of this Lagrangian 
depends on the sign of $\alpha$. If it is positive, the minimum occurs at $\varphi=0$. If it is negative, $\varphi=0$ is actually a maximum, and the minima occur at $\langle\varphi,\varphi\rangle=-\alpha/2\beta$.

In the former case, we can linearise about $\varphi=0$. The Lagrangian (in unrenormalised form) will be the complex Klein-Gordon Lagrangian with $m=\sqrt{2 \alpha}$ together with a perturbation proportional to $\phi^4$. In the latter case, however, there is a continuous family of minima: if $\varphi$ is a minimum, so is $\mathbf{R}\varphi$, where $\mathbf{R}$ is any rotation operator, and we can choose to linearise about any one of them. If, for instance,  we choose to linearise about $\varphi_0=(\sqrt{-\alpha/2\beta},0)$ then the Lagrangian density becomes
\be
\mc{L}=\sum_{a<N}\frac{1}{2}(\partial_\mu \varphi^{a} \partial^\mu \varphi^a + \sqrt{-2\alpha}(\varphi^{a})^2) + 
\frac{1}{2}(\partial_\mu \varphi^{N} \partial^\mu \varphi^N) + \mbox{(higher-order interaction terms)}.
\ee
Manifestly, the free part of the theory is only symmetric under $SO(N-1)$, not $SO(N)$. In fact, it should be clear that the free part of a theory will be symmetric under that subgroup of the internal symmetry group \emph{which leaves the expansion point invariant}. (So expansions around $\varphi=0$ lead to no reduction of symmetry, since that state is invariant under all symmetry operations).

Of course, significant work is necessary to rigorously (even in the particle-physics sense!) extend these results to quantum fields, and such work lies outside the scope of this paper. But the main results are just what we would expect from the semiclassical formalism:
\begin{itemize}
\item Non-linear theories can be analysed by a perturbative expansion around some particular state (usually the vacuum --- \iec lowest-energy --- state, corresponding to classical expansion around the minimum in the potential).
\item That perturbative analysis has the form of a free-particle theory, with higher-order terms in the perturbation showing up as interactions between particles.
\item Any symmetry of the full theory which leaves the expansion state and the positive-frequency subspace invariant is a symmetry of the particle theory. 
\end{itemize}

\section{Irreducibility and antimatter}\label{irreducibility}

Renormalisation gives us the real answer as to why it makes sense to treat a field with a reducible symmetry group as a collection of different fields. Any field with an $N$-dimensional internal space can be treated as $N$ interacting fields, of course --- but as long as those fields can be transformed into one another by a symmetry (whether internal or spacetime) the renormalised parameters of each field will remain the same as every other field. No such guarantee holds for reducible fields: in general the renormalised mass, charge and other such parameters will vary from irreducible component to irreducible component, even if the bare values of those parameters were identical between components. It is because of renormalisation, then, that particles can be identified with the \emph{irreducible} representations of the symmetry group, and because of renormalisation that it makes sense to require of a field that the symmetry group acts on it irreducibly.

But there is a complication. Linear field theories, as I have defined them, are represented by sections of a real vector bundle; the internal symmetries act on a real vector space via a real representation. As we have seen, though, the one-particle Hilbert space is to be identified with the (completion of) the positive frequency part of the complexification of the space of solutions, so that the internal symmetries act on the one-particle Hilbert space via the complexification of the real representation. So it becomes a rather urgent question to ask whether the complexification of the real-irreducible action of  the internal symmetry group is complex-irreducible. And here we find --- and this is the central observation of this paper --- that different choices of \mc{G} give very different results.

Suppose, for instance, that the theory in question is Klein-Gordon theory with a three-dimensional internal space, so that the internal symmetry group is $SO(3)$ and acts on \mc{V}($\simeq\mathrm{R}^3$) by the standard representation.  The complexification of this action turns out to be complex-irreducible, so the one-particle quantum theory of this field theory really is a \emph{one-particle} theory. The particle has a three-dimensional internal degree of freedom, and by a choice of basis in the internal space we could talk of (say) `red', `green' and `blue' particles, but no basis choice is to be preferred over another.\footnote{Despite the use of colour labels here, this theory is not really chromodynamics;  there the symmetry group is $SU(3)$, not $SO(3)$.}

And now contrast the complex Klein-Gordon theory. Here $\mc{V}\simeq \mathrm{R}^2$, and \mc{G} is $SO(2)\simeq U(1)$, which again acts in the standard way. The complexification $\mc{V}^C$ of \mc{V} is a two-\emph{complex}-dimensional space, and it is easy to see that the complexification of \mc{G} is \emph{not} complex-irreducible.

Why is this? Recall that the infinitesimal generator $\mathrm{J}$ of \mc{G} obeys $\mathrm{J}^2=-1$, and so its eigenvalues must be roots of -1 --- and since $\mathrm{J}$ is real, if $\lambda$ is its eigenvalue then so must be $\lambda^*$. So $\mathrm{J}$ has $\pm i$ for eigenvalues. In fact, it is easy to find the corresponding eigenvectors: we have
\be \mathrm{J}(1,\mp i)=\pm i(1,\mp i).
\ee
More directly: \mc{G} is Abelian and so its members must be represented by operators with common eigenvectors, so its irreducible representations must be one-dimensional.

So the one-particle quantum theory of the complex Klein-Gordon theory falls apart into two components: $\mc{H}_{1P}=\mc{H}^+_{1P}\oplus \mc{H}^-_{1P}$, and a rotation $R(\theta)\in \mc{G}$ acts on $\mc{H}^\pm_{1P}$ by multiplication by $\exp(\pm i \theta)$. This is a fundamentally different situation from the previous one: instead of a three-complex-dimensional internal space with no preferred direction, we have two different sorts of particle which transform in an essentially different manner under \mc{G}. That is, we have matter and antimatter.

What is more, quanta of the complex Klein-Gordon equation interact amongst one another via electromagnetic interactions, which are in turn tied (via the gauge principle) to the action of the $U(1)$ group on the quantum state. Since $U(1)$ acts oppositely on the particle and antiparticle subspaces, it follows that particles in these subspaces behave in a fundamentally different way when interacting (in familiar terms: they have opposite charges). So our choice to regard particle and antiparticle as fundamentally different is further motivated: they can be distinguished by their dynamical behaviour.

Indeed, we are now in a position to give a more satisfactory account of how the gauge principle is to be applied to nonrelativistic quantum physics. At the field-theoretic level, the $U(1)$ internal symmetry is on a par with all other internal symmetries, and the gauge principle can be applied to each of them in the same way. Quantizing the matter field (but treating the gauge field as classical, as in the Aharonov-Bohm effect) and restricting attention to the one-particle subspace yields a quantum particle theory with an internal symmetry which becomes local when corresponding changes are made to the classical gauge field. In some cases, this quantum particle theory will have matter and antimatter subspaces, each irreducible under the global internal symmetry, and the gauge symmetry will therefore act separately on each subspace. And in one special case --- that where the internal symmetry is $U(1)$ --- then there will be no internal degree of freedom, and the internal symmetry will act only by phase transformations: $\exp(-i q\theta)$ for matter, $\exp(+i q\theta)$ for antimatter. As such, the gauge transformations look like localisations of ``mere'' phase transformations: in reality, though, this is just because our attention is confined to the matter subspace of the one-particle space. In general, the global symmetry to which the gauge principle is to be applied is not the trivial, always-present
\be \label{simple}
\ket{\psi}\rightarrow \exp(-i\theta)\ket{\psi} 
\ee
but
\be \label{complicated}
\ket{\psi}\rightarrow\sum_{M=0}^{M=+\infty} \exp(-\op{J} M \theta) \Pi^M\ket{\psi}
\ee
where $\Pi^M$ projects out onto the $M$-particle subspace of Fock space and $\op{J}$ is the quantisation of the classical multiplication-by-i transformation. Only because of the behaviour of that transformation under complexification can~(\ref{complicated}) be written instead as 
\be
\ket{\psi}\rightarrow\sum_{N=-\infty}^{N=+\infty} \exp(-i N \theta) \Pi_N\ket{\psi}
\ee
where $\Pi_N$ projects onto that subspace of the Fock space where there are $N$ more particles than antiparticles, so that in the special case where $\Pi_N\ket{\psi}=\ket{\psi}$ (that is, where there is one particle and no anti-particle present) the transformation has the simple form~(\ref{simple}).

\section{Generalising the results}\label{generalise}

We have seen that the nature of the internal symmetries of quantum particles depends on whether the real-irreducible action of the internal symmetry group \mc{G} on the classical internal space \mc{V} is complex-irreducible when it is extended to $\mc{V}^C$, but so far we have looked at only two cases. Fortunately, there is a very elegant general theory here, to which we can appeal. I leave its full development to Appendix~A; here I will only sketch the results.

Firstly, suppose that \mc{G} is a real group acting complex-irreducibly on the complex vector space \mc{V}. As was stressed previously, any complex vector space is also a real vector space, so the complex action of \mc{G} is also a real action on that vector space; it is easy to see that it will be real-irreducible. Some --- but not all --- real irreducible representations can be obtained this way, and we will call those which are so obtained \emph{secretly complex}; the others we call \emph{honestly real}. (This is my own terminology, not something in general use.) A necessary and sufficient condition for a real representation to be secretly complex is that there exists a complex structure for \mc{V} (a real-linear map of \mc{V} to itself whose square is minus the identity)  which commutes with the action of \mc{G}.

(Conversely (though less crucially for our purposes), some \emph{complex} irreducible representations --- but not all --- can be obtained by complexifying a real-irreducible representation. We call those that are so obtained \emph{secretly real}; the others are (you guessed it) \emph{honestly complex}. A necessary and sufficient condition for a complex representation to be secretly real is that there exists a conjugation operation for \mc{V} (an antilinear map of \mc{V} to itself whose square is the identity) which commutes with the action of \mc{G}.)

The central result that we need (proved in Appendix A) is the
\begin{description}
\item[Complexification Theorem:] If \mc{V} is a real vector space and $\varphi$ is an irreducible representation of \mc{G} on \mc{V}, then:
\begin{enumerate}
\item[(a)]If $\varphi$ is honestly real, its complexification is irreducible.
\item[(b)]If $\varphi$ is secretly complex, its complexification is reducible: if $*$ is the natural conjugation map on the complexification $\mc{V}^\mc{C}$ of \mc{V},\footnote{\iec, if $(v,w)^*=(v,-w)$.} then there is a decomposition $\mc{V}^\mc{C}=\mc{V}^+\oplus \mc{V}^-$, where:
\begin{enumerate}
\item[(i)] $\mc{V}^+$ and $\mc{V}^-$ are conjugate: $(\mc{V}^\pm)^* =\mc{V}^\mp$;
\item[(ii)] the restrictions $\varphi^\pm$ of $\varphi$ to $\mc{V}^\pm$ are irreducible;
\item[(iii)] $\varphi^+$ and $\varphi_-$ are conjugate: $(\varphi^\pm(g)v)^*=\varphi^\mp(g)v^*$. 
\end{enumerate} \end{enumerate}
\end{description}

Put another way, the complexification of a real-irreducible representation of a group \mc{G} on a real vector space \mc{V} is either (i) complex-irreducible, or (ii) reduces into two irreducible conjugate actions, according to whether there (i) is not, or (ii) is, a linear transformation $\mathrm{J}$ on \mc{V} whose square is $-1$ and which commutes with \mc{G}.

For quantum fields, then, the situation is the following. 
\begin{itemize}
\item If a field is ``secretly complex'' (as is, for instance, the complex Klein-Gordon field) and has $N$ complex degrees of freedom, its one-particle quantum theory has a Hilbert space which decomposes into particle and antiparticle subspaces. Both particle and antiparticle have $N$ internal degrees of freedom; each carries an irreducible representation of the symmetry group.
\item If a field is ``honestly real'' and has $N$ real degrees of freedom, its one-particle quantum theory is acted on irreducibly by the internal symmetry group, and does not decompose into particle and antiparticle subspaces. The single particle has $N$ internal degrees of freedom.
\end{itemize}
What is the relation between the matter and antimatter subspaces? It might appear that the answer follows directly from the Complexification Theorem: isn't the representation of the symmetry group on one space the conjugate of its representation on the other space, so that the two spaces are naturally anti-isomorphic? This is in fact true for relativistic QFT, but the reasoning needs to be more subtle, for the following reason: as we have seen, the one-particle subspace is not simply the (completion in norm of) the complexification of the classical solution space, it is the positive-frequency subspace of that space. It is not prima facie guaranteed that this subspace should split naturally into anti-isomorphic matter and antimatter subspaces.

This can be illustrated by considering \emph{non}-relativistic quantum field theory. We can take the one-particle Schr\"{o}dinger equation and consider it as a field equation for a classical complex field: specifically,
\be
J \pbp{\psi}{t} - \frac{\nabla^2 \psi}{2m}=0.
\ee
(As usual, we are writing a complex field as a two-component real field.) The solution space $\mc{S}$ of this theory can be conveniently represented as the space of all functions of form
\be
\psi(x,t)=\int_{R^3} \dr{k} \left(C(k)(1+iJ)\e{-i(k\cdot x-\omega(k)t)} + C^*(k)(1-iJ)\e{-i(k\cdot x-\omega(k)t)}\right)
\ee
(where $k$ takes values in $R^3$, $\omega(k)=k^2/2m$, and $C(k)$ will need to satisfy appropriate boundary conditions: say, being the Fourier transform of a smooth function of compact support).
The norm-completion of its complexification $\mc{S}^C$ is the space of all functions of form
\be
\psi(x,t)=\int_{R^3} \dr{k} \left(C(k)(1+iJ)\e{-i(k\cdot x-\omega(k)t)} + D(k)(1-iJ)\e{-i(k\cdot x-\omega(k)t)}\right)
\ee
where $C(k)$ and $D(k)$ are arbitrary complex $L^2$ functions. $U(1)$ acts reducibly on this space; its irreducible subspaces are the spaces $\mc{S}^\pm$ of functions of form
\be
\psi^{\pm}(x,t)=\int_{R^3} \dr{k} \left(C(k)(1\pm iJ)\e{\mp i(k\cdot x-\omega(k)t)}\right)
\ee 
But $\mc{S}^+$ is also the norm-completion of the positive-frequency subspace of $\mc{S}^C$: that is, it is the one-particle subspace. So the antimatter subspace of the theory is entirely unphysical: in nonrelativistic quantum field theory there are particles but no antiparticles. (The physical content of this claim is that there are not both sorts of particle, of course; the actual labelling is just conventional.)

That this behaviour cannot occur in relativistic QFT is a consequence of the CPT theorem. To see why this is, we will need to extend our analysis to cover not just the small symmetries of a field theory --- spacetime and internal --- but also its large symmetries: space and time reflections, and charge conjugation.  This is the task of the next section.

\section{Large symmetries: field and particle}\label{CPT}

Before beginning a discussion of the large symmetries, a caveat is in order. Given the set of symmetries of a (classical or quantum) field theory, it is not actually a trivial matter to say which of those symmetries --- if any --- is ``the'' parity or time-reversal symmetry. ``Small'' spacetime symmetries like rotation or translation can (usually!) be operationally identified by considering physically rotating or translating the system, but (Alice in Wonderland notwithstanding) there is no straightforward way to perform a reflection, much less a time reversal, on a physical system. 

The defining features of a parity transformation, however, are that (i) the transformed field at spacetime point $(x,t)$ depends only on the untransformed field at $(-x,t)$; (ii) the transformation bears the appropriate algebraic relations to the other spacetime symmetries; (iii) the square of the transformation is (up to a phase factor) the identity. Similarly, a time reversal transformation has properties (ii) and (iii), and the 
transformed field at spacetime point $(x,t)$ depends only on the untransformed field at $(x,-t)$. And a parity-plus-time-reversal transformation has properties (ii) and (iii), and the transformed field at spacetime point $(x,t)$ depends only on the untransformed field at $(-x,-t)$.

With this in mind, let us make the following definitions. A classical field theory has
\begin{itemize}
\item \emph{parity symmetry} (\emph{time reversal symmetry, parity-time symmetry}) if there is a parity transformation (time reversal transformation, parity-plus-time-reversal transformation) that is a symmetry of the theory;
\item \emph{conjugation symmetry} if the internal symmetry group is secretly complex with complex structure \textbf{J} and there is a symmetry \textbf{C} such that $\mathbf{C}^2=1$ and $\mathbf{C}\mathbf{J}=-\mathbf{J}\mathbf{C}$. We will mostly be concerned with \emph{internal} conjugation symmetries: those where the transformed field at $(x,t)$ depends only on the untransformed field at the same point.
\end{itemize}
And a quantum field theory has
\begin{itemize}
\item \emph{P symmetry} (\emph{T symmetry, PT symmetry}) if there is a parity transformation (time reversal transformation, parity-plus-time-reversal transformation) of the one-particle subspace that is a symmetry of the theory \emph{and which takes particles to particles and anti-particles to anti-particles}.
\item \emph{CP symmetry} (\emph{CT symmetry, CPT symmetry}) if either (i) the theory has antiparticles and there is a parity transformation (time reversal transformation, parity-plus-time-reversal transformation) of the one-particle subspace that is a symmetry of the theory \emph{and which takes particles to anti-particles and anti-particles to particles}; or (ii) the theory does not have antiparticles and it has P symmetry (T symmetry, PT symmetry).
\item \emph{C symmetry} if either (i) the theory has antiparticles and there is an internal symmetry \textbf{C} of the one-particle subspace which takes particles to antiparticles and antiparticles to particles, or (ii) the theory does not have antiparticles.\footnote{We could get around this awkwardness by adopting the standard physicists' convention that some particles are their own antiparticles; however, for these purposes I find it clearer not to do so.}
\end{itemize}
These definitions of $C$, $P$ and $T$ and their products conform to standard usage in QFT; the question to be answered in this section is what their relations are to the symmetries of the classical field whose quantization produced the quantum field. Recall that a classical symmetry is a one-particle symmetry if it leaves invariant the vacuum state and the positive frequency subspace of the one-particle Hilbert space. We will assume that the vacuum is invariant under any of the symmetries we are currently considering, which leaves us with the question of whether or not any of them preserves the positive-frequency subspace. 

The answer is straightforward, in fact: parity and internal conjugation symmetries preserve it; time and parity-time symmetries swap the positive and negative frequency subspaces over.  This should be obvious for parity, time and parity-time symmetries, but at first sight it might appear that an internal conjugation symmetry also ought to interchange positive and negative frequency subspaces --- after all, doesn't the conjugation of $\exp(+i\omega t)$ equal $exp(-i\omega t)$?

This is confused in an important way, though. Recall that there are \emph{two} complex structures in play here: the complex structure in the classical complex field (which only some classical fields have), and the complex structure introduced when we complexify the space of classical solutions (which is present in all quantum theories).\footnote{There is a considerable history of discussions on this point; see~\citeN{saunderscomplexnumbers} and references therein.} With respect to that second complex structure, conjugation is real-linear (being the complexification of a linear map), and (if it is also internal) it leaves the positive-frequency subspace pointwise invariant.

Conversely, there is another relevant transformation: conjugation with respect to the quantum-mechanical complex structure, which we might call \emph{Q-conjugation} to distinguish it from the classical conjugation. \emph{This} conjugation operation interchanges positive and negative frequency subspaces, and so is not a symmetry of the one-particle Hilbert space; it is, however, clearly a symmetry of the complexified solution space. It follows that if we conjoin Q-conjugation with the time reversal operation, we will get a symmetry of the one-particle Hilbert space which is a time reversal symmetry (and similarly for parity-plus-time-reversal operations).

So: the one-particle Hilbert space has a parity, time-reversal, parity-time, or internal conjugation symmetry if and only if the classical theory does. To connect all this with $C,P,T$ and the like, however, requires us to consider what effects these symmetries have on the matter and antimatter subspaces. 

To answer this, let \textbf{J} be the complex structure for the classical field theory, and recall that particles and antiparticles are eigenstates of \textbf{J} with eigenvalue +1 (particles) or -1 (antiparticles). So as expected, an internal conjugation symmetry swaps particles and antiparticles.

Now, let \textbf{C}, \textbf{P}, \textbf{T} and \textbf{PT} be the real-linear operators that carry out the conjugation, parity, time-reversal and parity-time symmetry transformations on the classical fields.\footnote{This notation is confusing in one respect: \textbf{PT} is not necessarily the product of \textbf{P} and \textbf{T}, although it will be if \textbf{P} and \textbf{T} both exist.} If \textbf{X} is any of \textbf{P}, \textbf{T} and \textbf{PT}, then since \textbf{J} is a symmetry which commutes with all other internal symmetries, so will $\textbf{J}'=\mathbf{X}\mathbf{J}\mathbf{X}^{-1}$. The irreducibility of the internal symmetry group then entails that $\textbf{J}'=\pm\textbf{J}$. (To see this, complexify the solution space \mc{S} to $\mc{S}^\mc{C}$. The eigenvalues of \textbf{J} and $\mathbf{J}'$ are $\pm i$; since they commute with each other, they have joint eigenspaces. Let $\mc{S}_{++}$ be the eigenspace of vectors $v$ such that $\mathbf{J}v=\mathbf{J'}v=+i v$; let $\mc{S}_{+-}$ be the eigenspace of vectors $v$ such that $\mathrm{J}v=+iv$ and$\mathrm{J'}v=-i v$; define $\mc{S}_{-+}$ and $\mc{S}_{--}$. Then the sets $\{v+v^*:v\in \mc{S}_{++}\}$ and $\{v+v^*:v\in \mc{S}_{+-}\}$ are irreducible subspaces of \mc{S}. Since the internal symmetry group acts irreducibly on \mc{S}, either $\mc{S}_{++}=\emptyset$ or $\mc{S}_{+-}$ must be empty. Since $\mc{S}_{-+}=\mc{S}^*_{+-}$ and $\mc{S}_{--}=\mc{S}^*_{++}$, it follows that $\mathbf{J}'=\pm\mathbf{J}$. ) 

It follows that \textbf{P}, \textbf{T} and \textbf{PT} all either commute or anticommute with \textbf{J}; by definition, each anticommutes iff it is also a conjugation symmetry.
So a classical parity symmetry induces a $P$ or $CP$ symmetry according to whether or not it is a conjugation symmetry. Things are slightly more complex for \textbf{T} and \textbf{PT}: recall that the induced quantum symmetry from a time-reversal or parity-time symmetry is \textbf{T} or \textbf{PT} \emph{combined with Q-conjugation}. And Q-conjugation, of course, also swaps matter and antimatter. A classical time reversal symmetry, then, induces a $CT$ symmetry if it is \emph{not} a conjugation symmetry, a $T$ symmetry if it is; similarly for parity-time symmetries.

Since a real-linear transformation of a complex vector space is complex-(anti)linear if it (anti)commutes with the complex structure, parity, time and parity-time transformations are complex-antilinear if they are conjugate symmetries, and complex-linear if they are not. So we can summarise our results as follows:
\begin{itemize}
\item If a classical field theory has an antilinear internal symmetry, its quantization has $C$ symmetry.
\item If a classical field theory has a complex-linear parity symmetry, its quantization has $P$ symmetry; if the parity symmetry is complex-antilinear, its quantization has $CP$ symmetry.
\item If a classical field theory has a complex-antilinear time reversal symmetry, its quantization has $T$ symmetry; if the time reversal symmetry is complex-linear, its quantization has $CT$ symmetry.
\item If a classical field theory has a complex-antilinear parity-time symmetry that \emph{is} a conjugation symmetry, its quantization has $PT$ symmetry; if the parity-time symmetry is complex-linear, its quantization has $CPT$ symmetry.
\end{itemize}

The classical translation of the CPT theorem, then, is the requirement that classical fields have a complex-linear parity-time symmetry.

We are now in a position to answer the question left hanging at the end of the last section: what is the relation between the matter and antimatter sectors of the one-particle Hilbert space? The answer is that the irreducible actions of the symmetry group on the two sectors will be each other's conjugate, and so the two sectors will be naturally anti-isomorphic, provided that the theory has CPT symmetry, or (equivalently) that the classical field theory has a complex-linear parity-time symmetry. The non-relativistic Schro\"{o}dinger equation, interpreted as a classical field theory, has no such symmetry (it has a complex-linear parity symmetry and a complex-antilinear time symmetry); no wonder, then, that it has no antiparticles. But in relativistic quantum field theory, the CPT theorem tells us that all field theories have such a symmetry; hence, all relativistic quantum fields with internal $U(1)$ symmetry have naturally anti-isomorphic particle and antiparticle sectors.

\section{Conclusion}\label{conclusion}
 
Matter comes in particle and antiparticle form, and does so in some cases and not in others, because:
\begin{enumerate}
\item Particles are emergent phenomena, which emerge in domains where the underlying quantum field can be treated as approximately linear.
\item The wavefunctions of those particles obey the complexification of the linear dynamical equations governing the (linearisation of the) underlying field.
\item As such, the internal symmetry group of a particle is the complexification of the internal symmetry group of the underlying field.
\item As a matter of group theory, if an irreducible real representation of a group is complexified, it will either remain irreducible, or fall apart into two irreducible conjugate representations. In the first case, there will only be one species of particle; in the second case, there will be two species, and the CPT theorem ensures that the two species do not differ intrinsically (in terms of mass, charge, spin et al) but only relationally.
\item Both cases occur in nature.
\end{enumerate}
As for the gauge argument: it is really an argument whose natural home is classical field theory: there, any field theory with an rigid internal symmetry --- be it $U(1)$ or $SO(13)$ --- can be ``gauged'', turning the symmetry into a gauge symmetry and introducing a dynamical connection. When we quantize those gauge field theories whose symmetry group is honestly real, the result is that the symmetry group just acts on the internal space of the particles of the quantized field, in the way we would naturally expect. But if the internal symmetry group is secretly complex, then it acts reducibly on the particle and antiparticle sectors of the quantized field's internal space. In particular, if the internal symmetry group is $U(1)$ --- that is, if the classical field is complex --- then the group acts by phase rotation on each space of definite (particle-minus-antiparticle) number. In the particular case where one particle and no antiparticle are present, the action of $U(1)$, if the charge is $q$, is just ($\theta \rightarrow\exp(-iq\theta)$), creating the illusion that the global symmetry to be gauged is simply global phase rotations. But an illusion is all that it is. 

\section*{Acknowledgments}

I would like to thank Hilary Greaves and Simon Saunders for many useful discussions, and two anonymous referees for useful comments.

\section*{Appendix A: Complexification of real vector spaces}

(The results in the Appendix are fairly widely known but seem to have the status of folklore: detailed discussions at the appropriate level are hard to come by. The proofs, and some of the terminology, are my own, but I make no claim to originality.)

The basic theme of this appendix is that there are two ways to associate a real vector space with a complex one or vice versa: an easy way, and a hard way.

To begin with, suppose that \mc{V} is a real vector space; then a \emph{complex structure} \textbf{J} for \mc{V} is a (real-)linear map from \mc{V} to itself satisfying $\mathbf{J}^2=-1$. Complex structures let us turn real vector spaces into complex vector spaces. Since the only difference between the two is that a complex vector space needs to have a rule for multiplying vectors by complex scalars, all we need to do is define such a rule; and we do so by
\be
(\alpha+i\beta)v=\alpha v+ \beta\mathbf{J} v.
\ee
So: identifying a complex structure \textbf{J} on a real vector space $\mc{V}$ allows us to create a complex vector space $\mc{V}_\mathbf{J}$: indeed, $\mc{V}$ and $\mc{V}_\mathbf{J}$ are really the same space, just equipped with different structures. We will call this process \emph{remembering} \textbf{J}, for reasons which will become clear. However, finding a complex structure on a vector space is a non-automatic task. It may not even be possible (in particular, in the finite-dimensional case, only even-dimensional vector spaces possess complex structures); where it is possible, it will be highly non-unique.

On the other hand, there is a canonical way, given a real vector space $\mc{V}$, to associate a complex vector space with it. The process is called \emph{complexification}, and it has two steps. Firstly, we construct the direct sum of $\mc{V}$ with itself: $\mc{V}\oplus \mc{V}$. We then equip the resultant space with the following complex structure:
\be
\mathrm{J}(v,w)=(w,-v)
\ee
and define $\mc{V}^\mc{C}=(\mc{V}\oplus \mc{V})_\mathrm{J}$. Obviously, the complexification of $\mc{V}$ is a larger space than \mc{V} itself.

Going in the other direction, there is an extremely straightforward way to associate a real vector space with a complex one.  If \mc{W} is a complex vector space, it is already a real vector space: all we need to do is to restrict ourselves to multiplying vectors by \emph{real} scalars, and ``forget'' that we actually know how to multiply them by complex scalars too. This process, which we call \emph{forgetting}\footnote{The terminology here originates in category theory.}, generates a real vector space $\mc{W}^\mc{F}$ from the complex vector space $\mc{W}$: indeed, \mc{W} and $\mc{W}^\mc{F}$ are the \emph{same} space, just equipped with different structures. 

As for the hard way: If \mc{W} is a complex vector space, then a \emph{conjugation map} \textbf{C} for \mc{W} is an \emph{anti-linear} map of \mc{W} to itself satisfying $\mathbf{C}^2=+1$. (An antilinear map is real-linear and satisfies $\mathbf{C}(\alpha v)=\alpha^*\mathrm{C}v$).
Conjugation maps let us generate real vector spaces from complex ones. Specifically, consider the set of all vectors satisfying $\textbf{C}v=v$. This set is a real-linear subspace of the real vector space $\mc{W}^\mc{F}$, which we call the \textbf{C}-restriction of $\mc{W}$ and denote $\mc{W}_C$: clearly, it is a smaller space than $\mc{W}$.

Forgetting and remembering are inverses, in the following sense: if we start with a complex vector space $\mc{W}$, then the rule for multiplying vectors by $i$ defines a complex structure on $\mc{W}^\mc{F}$; remembering that structure gets us back to $\mc{W}$ again; and, if we start with a real vector space \mc{V}, remember a complex structure on \mc{V}, and then forget it again, we get back to \mc{V}. Similarly, there is a sense in which complexification and restriction are inverses. If we start with a real vector space and complexify it, there is a conjugation map defined on it via $\mathbf{C}(v,w)=(v,-w)$, and restriction via that conjugation map generates a real vector space isomorphic to the original one; and if we start with a complex vector space, restrict it via some conjugation map, and then complexify it, the resultant complex vector space is isomorphic to the original one.\footnote{As things stand, these isomorphism results are unimpressive, though, conveying no more than the fact that the vector space has a certain cardinality; they will gain more content shortly.} We call this conjugation map the \emph{natural} conjugation map on $\mc{V}^\mc{C}$, and usually write $\mathbf{C}v$ as $v^*$.

Now for group actions. Suppose that \mc{G} is a (real) Lie group; recall that an representation of \mc{G} on a real (resp. complex) vector space is a homomorphism of \mc{G} into the real (resp. complex) linear operators on that space. 

To begin with, let $\varphi$ be a representation of \mc{G} on a real vector space $\mc{V}$. Then the complexification $\varphi^\mc{C}$ of $\varphi$ is a representation of \mc{G} on the complexification $\mc{V}^\mc{C}$ of \mc{V}, defined by
\be
\varphi(g)^\mc{C}(v,w)=(\varphi(g)v,\varphi(g)w).
\ee
It is easy to see that $\varphi^\mc{C}$ is indeed an action of \mc{G}.

Conversely, suppose that $\varphi$ is a representation of \mc{G} on a complex vector space $\mc{W}$. Then trivially, $\varphi$ is also a representation of \mc{G} on the real vector space $\mc{W}^\mc{F}$. To distinguish this real representation from the (essentially identical) complex representation, we write it as $\varphi^\mc{F}$.

So, the ``easy'' ways of associating real and complex vector spaces with one another have associated ``easy'' ways of relating real and complex representations with one another. Similarly, the ``hard'' ways have associated ``hard'' (\iec, non-automatic) ways of relating representations.

To begin with the remembering operation: suppose that \mc{V} is a real vector space on which \mc{G} acts by some representation $\varphi$. If there is a complex structure \textbf{J} on \mc{V} such that $\varphi(g)$ commutes with $\mathbf{J}$ for all $g\in \mc{G}$, we say that $\varphi$ is \emph{secretly complex}; if not, $\varphi$ is \emph{honestly real}. (This slightly whimsical terminology is mine: in the literature it is  more common just to say ``real'' and ``complex'', but where representations of a group on a complex vector space can be real and vice versa, confusion beckons.) 

If $\varphi$ is secretly complex, it is easy to see that it is also a representation of \mc{G} on the complex vector space $\mc{V}_\mathrm{J}$. 

Conversly: suppose that \mc{W} is a complex vector space on which \mc{G} acts by some representation $\varphi$. If there is a conjugation map \textbf{C} on \mc{W} such that $\varphi(g)$ commutes with $\mathbf{C}$ for all $g\in \mc{G}$, we say that $\varphi$ is \emph{secretly real}; if not, $\varphi$ is \emph{honestly complex}.\footnote{Since any two conjugate maps are related by some linear isomorphism, a complex representation is secretly real iff its image under the adjoint action of some \emph{fixed} conjugation map is \emph{equivalent} to the original representation} If $\varphi$ is secretly real, it is easy to see that its restriction $\varphi_\mathbf{C}$ to $\mc{W}_\mathbf{C}$ is a real representation of \mc{G}. Furthermore, if $\varphi$ is a real representation of \mc{G} on \mc{V} and \textbf{C} is the natural conjugation map on $\mc{V}^\mc{C}$, then $(\varphi^\mc{C})_\mathrm{C}$ is equivalent to $\varphi$.

It is easy to see that a complex representation is secretly real iff it is the complexification of a real representation, and that a real representation is secretly complex iff it is obtained from a complex representation by forgetting the complex structure.

We can now prove our goal.
\begin{description}
\item[Complexification Theorem:] If \mc{V} is a real vector space and $\varphi$ is an irreducible representation of \mc{G} on \mc{V}, then:
\begin{enumerate}
\item[(a)]If $\varphi$ is honestly real, its complexification is irreducible.
\item[(b)]If $\varphi$ is secretly complex, its complexification is reducible: if $*$ is the natural conjugation map on $\mc{V}^\mc{C}$, then there is a decomposition $\mc{V}^\mc{C}=\mc{V}^+\oplus \mc{V}^-$, where:
\begin{enumerate}
\item[(i)] $\mc{V}^+$ and $\mc{V}_-$ are conjugate: $(\mc{V}^\pm)^* =\mc{V}^\mp$;
\item[(ii)] the restrictions $\varphi^\pm$ of $\varphi$ to $\mc{V}^\pm$ are irreducible;
\item[(iii)] $\varphi^+$ and $\varphi_-$ are conjugate: $(\varphi^\pm(g)v)^*=\varphi^\mp(g)v^*$. 
\end{enumerate} \end{enumerate}

\end{description}
\noindent \textbf{Proof:} Let $\mc{V}_0$ be an irreducible subspace of \mc{V}. $\mc{V}^*_0$ must also be invariant under the action of \mc{G}, for \be\varphi(g)(v,w)^*=\varphi(g)(v,-w)=(\varphi(g)v,-\varphi(g)w)=(\varphi(g)(v,w))^*.\ee
There are now two possibilities:
\begin{enumerate}
\item $\mc{V}_0$ has non-trivial overlap with $\mc{V}^*_0$. Then there is a vector $(v,w)$ such that both $(v,w)$ and $(v,-w)$ are in $\mc{V}_0$, and so (by linearity) is $(v,0)$. So there is a nontrivial real subspace $\mc{X}$ of $(\mc{V}_0)$. Since $\varphi^\mc{C}(g)(v,0)=(\varphi(g)v,0)$, \mc{X} is isomorphic to a $\varphi-$invariant subspace of $\mc{V}$ under the isomorphism $v \rightarrow (v,0)$. But by assumption $\varphi$ is irreducible, so the only such subspace is \mc{V} itself. Hence $\mc{V}_0=\mc{V}^\mc{C}$.
\item $\mc{V}_0$ overlaps trivially with its own conjugate. Then $\mc{V}_0\oplus \mc{V}_0^*$ is well-defined, and contains a real subspace. By the argument above, that subspace must be invariant under the action of \mc{G}, and so must be \mc{V} itself; hence, $\mc{V}_0\oplus \mc{V}_0^*=\mc{V}^\mc{C}$.
\end{enumerate}

So we have shown that any irreducible representation, once complexified, is either irreducible or factors into two irreducible representations conjugate to one another. It remains to be shown that the latter case occurs iff the representation is secretly complex. We do this via the isomorphism between $\varphi$ and the restriction of $\varphi^\mc{C}$ to the conjugation-invariant subspace of $\mc{V}^\mc{C}$, which allows us to identify the two spaces. 

If $\varphi^\mc{C}$ is reducible, we can write any vector $w$ in $\mc{V}$ uniquely as $w=w^++w^-$, where $w^\pm$ is the projection of $w$ onto $\mc{V}^\pm$. Clearly $w^-=(w^+)^*$; conversely, if $v\in \mc{V}^+$ then $v+v^*$ is a vector in \mc{V}. Furthermore (again via the isomorphism) ($\varphi(g)(v+v^*)=\varphi^+(g)v+\varphi^-(g)v^*$).  We now define a complex structure \textbf{J} on \mc{V} by $\mathbf{J}(v+v^*)=iv+(iv)^*$. It is easy to check that this structure is invariant under $\varphi$; hence, $\varphi$ is secretly complex.

Conversely, suppose that \textbf{J} is a complex structure on \mc{V}. \textbf{J} can be extended by complex linearity to $\mc{V}^\mc{C}$, and it is easy to verify that the subspaces $\mc{V}^\pm=\{(v\pm iJ v):v\in \mc{V}\}$ are non-identical and invariant under $\varphi^\mc{C}$; hence, $\varphi^\mc{C}$ is reducible. $\Box$

\section*{Appendix B: The Standard Model}\label{standardmodel}

How does all this work out for actual quantum field theories? In this appendix I will briefly illustrate how it applies to the Standard Model of particle physics. By necessity, the account is relatively technical, and will presume some familiarity with gauge theories, to about the level presented in~\citeN{peskinschroeder}.

Basically (at least up to the standard model), pretty much all field theories used in particle physics fall into one of four types:
\begin{enumerate}
\item Scalar fields obeying the Klein-Gordon equation.
\item Weyl spinor fields.
\item Majorana spinor fields.
\item Real vector fields.
\end{enumerate}
(By the spin-statistics theorem, of course, the vector and scalar fields must be quantized as bosonic fields, and the spinor fields as fermionic fields.) Any of these spaces may be equipped with additional internal degrees of freedom by replacing scalars, spinors or vectors with \mc{V}-valued scalars, spinors or vectors (or, equivalently, by replacing the existing internal space by its tensor product with \mc{V}). It will be convenient to call \mc{V} the \emph{purely internal space}: in the case of scalar fields, the purely internal space just is the internal space, but in the other cases it is smaller.

Dirac spinors are left off this list for two reasons. Firstly, in the Standard Model, Dirac spinors are normally considered to be Weyl spinors connected by mass interaction terms that emerge from symmetry breaking; more importantly from this paper's perspective, a Dirac spinor field is just a Majorana field with an internal complex degree of freedom.

In the absence of interactions, the internal symmetry groups can be written down explicitly. If the internal space has $N$ (real) dimensions, then it is $SO(N)$ in the case of scalar, Majorana and vector fields; in the case of Weyl spinors, it is $SU(N)$. In fact, the Standard Model (at the classical level) is generally constructed by starting with some set of scalars and spinors (representing matter), equipping them with internal degrees of freedom, and then ``gauging'' the resultant internal symmetry by adding a connection. That connection, in turn, is represented by a vector field taking values in the Lie algebra of the gauge group. The internal symmetries then act on that vector field by the adjoint action, with the consequence that the internal symmetry group of the vector field is generally smaller than SO(N). If the gauge group is $SU(2)$, for instance, then the vector field representing the gauge connection takes values in a three-real-dimensional space (the Lie algebra $su(2)$; since (at least locally) $SU(2)\simeq SO(3)$, in this case the internal symmetry group in the interacting case is the same as in the free case. But if the group is $SU(3)$, then the vector field takes values in $su(3)$, which is eight-dimensional. $SO(8)$ is a considerably larger group than $SU(3)$.

If neutrino mass is ignored, the simplest form of the standard model has, as internal symmetry group, $SU(3)\times SU(2)\times U(1)$. Its components are:
\begin{description}
\item[Left-handed lepton fields:] Three left-handed Weyl spinor fields each with two purely internal real degrees of freedom. The internal space of the fields is then a four-complex-dimensional complex vector space, which we can write as $\mathrm{C}^2\otimes\mathrm{C}^2$. The spacetime symmetries act only on the first term in the product, $SU(2)$ acts on the second term by the direct action,  $U(1)$ acts by the direct action multiplied by a charge term ($\theta \cdot\psi-\exp(i q \theta) \psi$), and  $SU(3)$ acts trivially.
\item[Left-handed quark fields:] Three left-handed Weyl spinor fields each with six purely internal real degrees of freedom, so that the internal space is twelve-complex-dimensional. It can be factorised as $\mathrm{C}^2\otimes\mathrm{C}^2\otimes\mathrm{C}^3$, with the spacetime symmetries acting only on the first term in the product, with $SU(2)$ acting on the second term by the direct action, with $SU(3)$ acting on the third term by the direct action, and with $U(1)$ acting by the direct action multiplied by a charge term.
\item[Right-handed lepton fields:] Three right-handed Weyl spinor fields each with no purely internal complex degrees of freedom. $SU(2)$ and $SU(3)$ act trivially; $U(1)$ acts by the direct action multiplied by a charge term.
\item[Right-handed quark fields:] Three right-handed Weyl spinor fields each with three purely internal real degrees of freedom, so that the internal space can be written as $\mathrm{C}^2\times\mathrm{C}^3$, with the spacetime symmetries acting only on the first term in the product, with $SU(3)$ acting on the second term by the direct action, with $U(1)$ acting by the direct action multiplied by a charge term, and with $SU(2)$ acting trivially. 
\item[Gauge fields:] Three massless vector field with the Lie algebras $su(2)$,$su(3)$ and $u(1)$ as their purely internal spaces, with $U(1)$, $SU(2)$ and $SU(3)$ acting by the adjoint actions on their respective Lie algebras. (The adjoint action of $U(1)$ on $u(1)$ is trivial, of course, since $U(1)$ is Abelian).
\item[Higgs field:] A scalar field with two internal complex degrees of freedom, on which $SU(2)$ and $U(1)$ act by the direct action (multiplied by a charge term, in the case of $U(1)$) and $SU(3)$ acts trivially.
\end{description}
Note that if all these terms are taken to be independent linear fields, the internal symmetry group is considerably bigger: we get $SU(2)\times SU(3)\times U(1)$ by insisting that the full interacting theory is symmetric under the action of the group. 

Let us start by assuming that the vacuum is invariant under the internal symmetry group. If this were true, then the internal symmetries of the linearised theory --- and so of the one-particle subspaces --- would again be $SU(2)\times SU(3)\times U(1)$. The direct action of each of the three groups is secretly complex; the adjoint action of each group is secretly real. So we predict that
\begin{enumerate}
\item The left-handed quark fields each have a secretly complex symmetry group, $SU(2)\times SU(3)\times U(1)$. So we expect their one-particle subspaces to break into quark and antiquark components. Since Weyl spinors are $CP$-invariant but not $P$-invariant, it causes confusion to refer to the antiquarks here as left-handed: an anti(left-handed quark) is a right-handed antiquark. So the particles are left-handed quarks and right-handed antiquarks, each with a six-dimensional purely internal space. Each particle, of course, is massless and obeys the Weyl equation (in left- or right-handed form, as appropriate).
\item Similarly, the symmetry groups of the other quark and lepton fields are secretly complex. So we expect right-handed quarks and left-handed anti-quarks with a three-dimensional purely internal space, left-handed leptons and right-handed antileptons with a two-dimensional purely internal space, and right-handed leptons and left-handed antileptons with no purely internal space.
\item The symmetry group of the Higgs boson is also secretly complex, so we expect Higgs bosons and anti-Higgs bosons, each with a two-dimensional internal space.
\item The symmetry group of the gauge fields are honestly real, so there are no gauge bosons and gauge antibosons, just three vector bosons with, respectively, a three-dimensional, eight-dimensional and trivial purely internal space.
\end{enumerate}
Matters are complicated only slightly by the fact that the theory also has $SU(2)\times SU(3)\times U(1)$ as a \emph{gauge} symmetry. The details of how to handle this can get messy (and are well beyond the scope of this paper), but in essence, the result is to remove twelve degrees of freedom from the gauge fields. A massless classical vector field appears to have four degrees of freedom, but one of the dynamical equations is actually a constraint equation (corresponding to Gauss's Law, in fact), which reduces it to three. A massless vector field without purely internal degrees of freedom, however, only has two degrees of freedom, so the gauge bosons have six, sixteen, and two. The other twelve degrees of freedom are absorbed by gauge symmetry.

But it is not actually true that the vacuum of the Standard Model is invariant under the full internal symmetry group. At sufficiently high energies it can be treated as such, and the description above is expected to be correct. But at lower energies we need to allow for the spontaneous symmetry breaking induced by this vacuum non-invariance. In fact, the vacuum is invariant under the action of $SU(3)$ but not under $SU(2)\times U(1)$. The simplest model for this symmetry breaking introduces a potential term $V(\phi)$ for the Higgs field which (i) is a function of $\langle \phi,\phi\rangle$ and (ii) has a nonzero minimum --- at $\phi_0$, say. Choosing particular coordinates for the internal space of the Higgs boson, we can arbitrarily take the particular vacuum around which we expand to be
\be
\phi=\vct{\phi_0}{0}.
\ee
A basis for $su(2)\times u(1)$ is the set $\{1,\sigma_x,\sigma_y,\sigma_z\}$, and it is easy to see that $\phi$ is invariant under the action of $1-q\sigma z$ and its multiples (where $q$ is the charge of the Higgs field) but not under any other element of the Lie algebra. So the $SU(2)\times U(1)$ symmetry is broken down to $U(1)$.

The symmetry breaking also introduces quadratic terms coupling left and right handed leptons and quarks together. Being quadratic, these terms show up as parts of the free-particle theory and not as interactions: they act as mass terms, coupling left and right handed Weyl fermions together into Dirac fermions. In the simplest form of symmetry breaking, the mass terms arise from terms like $\bar\psi_L \phi \psi_R$, where $\psi_L$  and $\psi_R$ are left and right handed Weyl spinors and $\phi$ is the Higgs scalar. If the lowest-energy value of the Higgs field were at $\phi=0$, this would be a cubic term corresponding to a three-particle interaction; since it is at $\phi_0 \neq 0$, we get the particle spectrum by expanding $\phi$ as $\phi=\phi_0+\delta \phi$, leading to a quadratic term $\bar\psi_L \phi_0 \psi_R$.

Looking again at the various fields of the standard model, we find that:
\begin{itemize}
\item The left-handed lepton is acted on reducibly by $U(1)$, so it ought to correspond to two distinct particles. One of those particles couples to the right-handed lepton to form a Dirac spinor; the other remains a Weyl spinor. Since both theories have a $U(1)$ symmetry, both can be found in both antimatter and matter forms. In fact, the Dirac spinors are the electron, mu meson, and tau meson; the Weyl spinors are the neutrinos.
\item Similarly, the left handed quarks are acted on reducibly by $U(1)\times SU(3)$, and break into pieces each of which has an internal space on which $SU(3)$ acts irreducibly. They also couple to the right-handed quarks to form Dirac spinors; each, having $U(1)$ as an internal symmetry, comes in antimatter and matter forms.
\item The gauge boson of $SU(3)$ --- the gluon --- is unaffected by the symmetry breaking. The generator, $s\equiv 1-q\sigma z$, of the $U(1)$ residual symmetry acts reducibly on the pair of vector bosons with joint Lie algebra $su(2)\times su(1)$. In fact, we can see that it has a two-dimensional irreducible subspace spanned by $\sigma_x$ and $\sigma_y$, and that it acts trivially on $1$ and $\sigma_z$. At the classical level, then, the linearized field equations include a vector field with a two-dimensional purely internal space on which $U(1)$ acts (that is: a complex vector field), and two vector fields without purely internal spaces. Quantizing, we would expect to find two vector bosons which do not come in matter and antimatter forms, and one which does. This is indeed what we find: the former two bosons are the photon and the $Z_0$ particle; the latter has the $Z_+$ and the $Z_-$ as its particle and its antiparticle.
\item The Higgs scalar's behaviour on symmetry breaking is complicated by gauge freedom. In fact, three of its four degrees of freedom (corresponding to perturbations of the \emph{direction} of the Higgs field in internal space around the vacuum direction) are absorbed entirely by gauge freedom and simply contribute mass terms to the $W$ and $Z$ bosons. The fourth degree of freedom (corresponding to perturbations of the \emph{magnitude} of the Higgs field) remains as a particle (the Higgs boson, still unobserved at time of writing); since the residual $U(1)$ symmetry acts trivially on such perturbations, the Higgs boson does not come in particle and antiparticle forms.
\end{itemize}
 
For completeness, I should note one last wrinkle. At long distances, the interactions of the $SU(3)$ gauge field with the quarks become so strong that the particle approximation breaks down entirely. A different particle approximation --- in terms of protons, neutrons, and other states which transform under $SU(3)$ as singlets --- takes over, but the details of this process lie far beyond this paper, and indeed cannot, at time of writing, be analysed quantitatively without the aid of numerical simulations.

\end{document}